# A Self-Authentication and Deniable Efficient Group Key Agreement Protocol for VANET

**Mu Han, Lei Hua, Wenshan Liu, Shidian Ma**

**Jiangsu University, China**

Mu Han[a], Lei Hua[a], Shidian Ma[b,*]

[a] School of Computer Science and communication Engineering, Jiangsu University, 212013 Zhenjiang, China

[b]School of Automotive Engineering Research Institute , Jiangsu University, 212013 Zhenjiang, China

**Abstract**

With the rapid development of vehicular ad hoc Network (VANET), it is gaining significant popularity and receiving increasing attentions from academics and industry in security and efficiency. To address security and efficiency issues, a self-authentication and deniable efficient group key agreement protocol is proposed in this paper. This scheme establishes a group between road-side unit (RSU) and vehicles by using self-authentication without certification authority, and enhances certification efficiency by using group key (GK) transmission method. At the same time, to avoid the attacker to attack the legal vehicle by RSU, we adopt deniable group key agreement method to negotiation session key (sk) and use it to transmit GK between RSU. In addition, vehicles not only broadcast messages to other vehicles, but also communicate with other members in the same group. So group communication is necessary in VANET. Finally, the security and performance analysis show that our scheme is security, meanwhile the verification delay, transmission overheard and message delay are more efficient than other related schemes in authentication, transmission and communication.

*Keywords:* Vehicular ad hoc network (VANET), Deniable, Group communication, Key negotiation

*

*Corresponding author. Tel.: +86 182 6195 6301; E-mail address: masd@ujs.edu.cn (S. Ma)

# 1 Introduction

VANET is essentially a branch of Mobile Ad hoc Networks (VANETs) and have many prominent features, such as Vehicle moving at high speed, rapidly changing topology, short interaction time between nodes and so on. In addition, the nodes of VANET compose two parts, one is vehicles equipped with on-board unit (OBU) and another is road-side unit (RSU), vehicles communicate with each other as well as RSU through open wireless channel[1]. Due to the peculiar attribute of VANET, it can provide some security services including traffic information, traffic safety warning and infotainment dissemination for drivers and passengers. However, VANET also faces many challenges, such as security and privacy. Therefore, VANET need to meet some security requirements to resist security threats, as follows:

- Message integrity and authentication: Vehicles could verity that a message is indeed sent and signed by another vehicle without being modified by anyone
- Privacy: Vehicle's real identity should not be linked to any message, and other vehicles or RSU cannot reveal any vehicle's real identity by analyzing multiple messages sent of it.
- Confidentiality: The message sent to dedicated vehicles should be readable only by the dedicated receiver, other vehicles should be unable to decrypt the message.
- Non-repudiation: To avoid the sender of a message denied that he had sent the message, trust authority (TA) could obtain a vehicle's real identity and related this message to the sender.
- Forward-security and backward-security: The vehicle which leaves the communication group should not acquire the new group key and could not communicate with original group members. In addition, a new group member should be unable to learn the previous group key.

So, we design a self-authentication and deniable efficient group key agreement protocol to meet above security requirements.

The remainder of the paper is organized as follows: Some related work is given in Section 2. Section 3 presents the system model, bilinear maps and hard problem. Section 4 proposed our scheme. Section 5 and Section 6 analyses security and performance of the proposed protocol. Finally, Section 7 concludes the paper.

## 2 Related work

With the development of VANET, the communication technology in it is being gradually improved and has been focused on two aspects. One is the communication between vehicles and RSUs, and another is the communication between vehicles. Because of the short communication range and the vehicle high speed moving, the efficiency and security of communication in VANET become extremely significant. To address the above problems,[1] and [2] produced the public and private keys by using PKI mechanism, and utilized digital signature certification to guarantee the validity of the vehicles. But the overhead of the signature, verification and transmission will increase rapidly along with the increasing of nodes. The exorbitant cost of transmission and calculation make the scheme more and more cannot be adopted. Aiming at the problem of communication efficiency, [3]and[4] adopted identity-based cryptography, not only reduced the computation process of the public key certificate, but also alleviated the computation and transmission overhead. Raya and Hubaux [5] proposed a batch verification scheme. In this scheme, the computation efficiency can be significantly improved, but it has the problem of privacy. Later, [6] introduced "an efficient identity-based batch verification scheme", in which the efficiency of the vehicle certification can be improved. However, both[5] and [6] have privacy issues. For the privacy, [7] suggested that every vehicle should be pre-loaded with a large number of anonymous public and private keys, and the corresponding public key certificates. It avoids being tracked to a certain extent by using this method, but it will waste a lot of time on checking the list of revoked certificates. [8], [9] by pre-loading the large pseudonym to achieve privacy protection.[10] proposed an efficient vehicle group forming technique, vehicles establish a reliable group to enhance the security and privacy protection by using trusted communication equipment (TPM, Platform Module Trusted). [11] proposed using two kinds of top-level mechanisms of PKI system, which are information signature and group signature. In addition, it also adopted batch verification method. So, this scheme not only met the requirements of privacy protection, but also reduced the computation overhead.

## 3    Preliminaries

### 3.1 Network Model

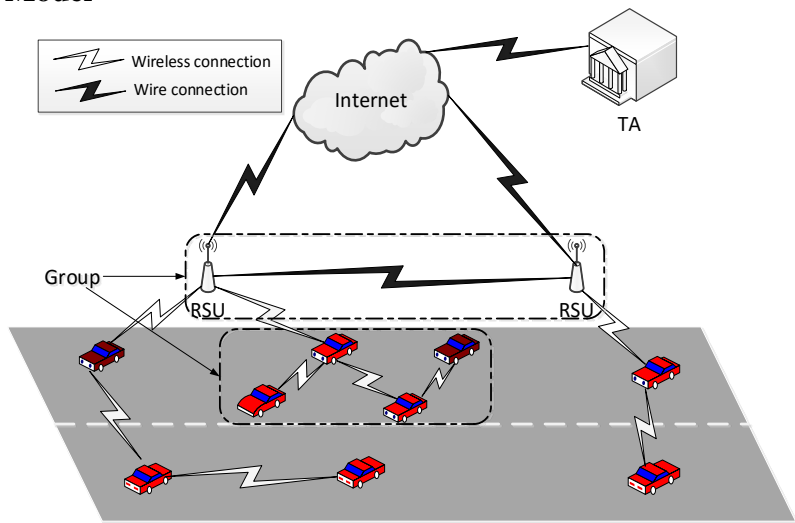

**Fig. 1.** System model

As shown in Fig. 1, the network model of VANET consists of three entities: The Trust Authority (TA), RSU at the road side and OBU equipped on mobile vehicles.

- TA: TA is a trusted certification center of the whole network. It responses to register and manages all nodes in the VANET and exposes the true identity of the valid vehicles and releases the information of the revoked vehicles. In this paper, TA allocate the certificate parameters to RSU and vehicles. As usual, TA is assumed powered with sufficient storage capability and infeasible for any adversary to compromise.

- RSU: RSU are densely distributed in the roadside. They connect with TA by wired links while using wireless links to vehicles. In this protocol, RSU is used to authenticate the validity of the vehicles and negotiate with the vehicles to form the group. In addition, it help TA liable of the vehicles which have illegal act. The adjacent RSU know each other's public key and can communicate with them.

- Vehicles: Vehicles are equipped with OBU, which periodically broadcast traffic related status information to improve the road environment, traffic safety, and infotainment dissemination for drivers and passengers[12].The communication among them is based on the DSRC protocol.

### 3.2 Bilinear Maps and Hard Problem

1) Bilinear Maps

Let $G_1$ be a cyclic additive group and $G_2$ is a cyclic multiplicative group. Both group $G_1$ and $G_2$ have the same prime order $q$ ( $k$ bites, $k$ is the safety parameters of the system). Let $P$ be a generator of $G_1$, $aP$ express $P$ self-increase $a \in Z_q^*$ times. A mapping $e: G_1 \times G_1 \to G_2$ is called bilinear mapping if it satisfies the following properties.

- Bilinearity: $e(aP, bQ) = e(P, Q)^{ab}$ for all $P, Q \in G_1 \quad a, b \in Z_q^*$.

- Nondegeneracy: $e(P, P) \neq 1$

- Computability: There is an efficient algorithm to compute $e(P, Q)$, for any. $P, Q \in G_1$

2) Computational Diffie-Hellman problem in $G_1$ (CDH Problem)

Let $P$ be the generator of $G_1$, for all $a, b \in Z_q^*$, give $(P, aP, bP)$, output $abP$ by the probabilistic polynomial time algorithm $A$. The probability of success is defined as

$$Succ_{A,G_1}^{CDH} = \Pr[A(P, aP, bP) = abp : a, b \in {}_R Z_q^*]$$

CDH Assumption: $Succ_{A,G_1}^{CDH}$ is a negligible value for all the PPT algorithm $A$.

3) Decide Diffie-Hellman problem in $G$ (DDH Problem)

Let $G$ be a special cyclic multiplicative group, $P$ is the safety parameter, and it satisfies $p = 2q + 1$. Let $g$ be the generator of the cyclic group $G' = <g>$, $G'$ is quadratic residue class of model $P$. Define the function $f : f(x) = \begin{cases} x & \text{if } x < q \\ p-x & \text{if } q < x < p \end{cases}$. Through function $f$, define $G$: $G = \{f(g^i) | i \in Z_q\}$. Define the exponent arithmetic in $G : a^b := f(a^b \bmod p)$, $a, b \in G$. The DDH problem in $G$ is that for $(g, g^x, g^y, g^r)$, $x, y, r \in_R G$, exist a PPT algorithm $A$ which output is $0/1$, output $1$ when $r = xy$; Otherwise, output $0$. The advantage of $A$ solve the DDH problem in $G$ define as follows:

$$ADv_{A,G}^{DDH} = |\Pr[A((g, g^x, g^y, g^{xy})) = 1] - \Pr[A((g, g^x, g^y, g^{xy})) = 1]| : \quad x, y, r \in_R G$$

DDH Assumption: $ADv_{A,G}^{DDH}$ is a negligible value for all the PPT algorithm $A$ which output is $0/1$.

**4. Proposed Scheme**

In this section, we describe our scheme with the following process: system initialization, the deniable group key negotiation of RSU, the authentication between RSU and Vehicles, the negotiation and update of the group key between RSU and Vehicles, the communication among the group. The notations used throughout the paper are listed in Table 1.

4.1 System Initialization

- Given the parameters $(G_1, G_2, P, q, e, G, p, g)$, which satisfy the description in Section 3.2, TA initializes the system by the following steps:

  1) TA chooses a random number $\psi_{TA} \in Z_q^*$ as its private key $SK_{TA}$ and compute the public key $PK_{TA} = g^{\psi_{TA}}$;

  2) TA chooses two cryptographic hash functions: $H_1: \{0,1\}^* \to G_1, h: \{0,1\}^* \to Z_q^*$;

  3) TA chooses a security symmetric cryptographic $E_k(\cdot)$, and then TA publishes the system parameters, which include $(G_1, G_2, P, q, e, G, p, g, PK_{TA}, H_1(\cdot), h(\cdot), E_K())$ and downloads these parameters into RSU and Vehicles.

- RSU require to download the system parameters by TA before being installed to the appropriate location. TA distributes a true identity $TID_{RSU_i}$ to $RSU_i$ and chooses a random $\xi_i \in Z_q^*$ as its private key $SK_{RSU_i}$ and computes the public key $PK_{RSU_i} = g^{\xi_i}$, the certification parameters $Q_{RSU_i} = H_1(TID_{RSU_i})$ and $s_{RSU_i} = \psi_{TA} Q_{RSU_i}$. Then download the public key, the private key and the certification parameters into $RSU_i$. At the same time, download other $RSU_i$'s public keys which is located nearby regions.

**TABLE 1** NOTATIONS AND DESCRIPTIONS

| Notations | Descriptions |
| --- | --- |
| $V_i, TID_{V_i}, FID_{V_i}$ | Vehicle $V_i$, true name, pseudonym |
| $R, TID_{RSU_i}$ | $RSU$, true name |
| m | message |
| $Q_U, S_U$ | The certification parameters of the node $U$ |
| TS | Time stamp |
| U | The node $U$, vehicle or $RSU$ |
| $RSU_{i\pm 1}$ | The $RSU$ nearby $RSU_i$ |
| $(PK_U, SK_U)$ | The public and private key of the node $U$ |
| $\sigma_{SK_U}(\cdot)$ | The signature of the node $U$ |
| GK | Group Key between $RSU$ and vehicles |
| $VVK_{i,j}$ | The shared key among vehicle $V_i$ and $V_j$ |
| $HMAC_k(\cdot)$ | The message authentication by symmetric key $k$ |
| $E_k(\cdot)$ | Encrypt by key $k$ |
| pid | The set of all users' $TID$ who participate in the process |
| sk | The session key in Group of $RSU$ |

- Vehicles also require to download the system parameters by TA before they used. TA distributes a true identity $TID_{V_i}$ to vehicle $V_i$, then TA computes the certification parameters $Q_{V_i} = H_1(TID_{V_i})$, $s_{V_i} = \psi_{TA} Q_{V_i}$ and downloads them into it. In order to ensure vehicles could not traced by the malicious nodes, when vehicles enter a new range of RSU, it will motivate the key generator to generate the private key $SK_{V_i} = \alpha_i, \alpha_i \in Z_q^*$, the public key $PK_{V_i} = g^{\alpha_i}$ and pseudonym $FID_{V_i} = TID_{V_i} \oplus H_1(\alpha_i * PK_{TA})$

## 4.2 Deniable group key negotiation of RSU

We adopted two-round deniable group key agreement protocol[13], which is used to establish a confidential channel for communications. Simultaneously, it allows participants to deny that they have ever participated in group key agreement. Firstly, we contract a group (it include all RSU) and then generate the session key (sk) between RSU by using deniable group key agreement, which is the preparation work for the group key transmission mechanism. The purposed is to prevent the attacker to track the legitimate vehicles from RSU and ensures the security of group key transmission. The deniable group key negotiation of RSU as follows:

Step 1: $RSU_j$ random select $x_j, r_j, t_j \in Z_q$ compute $X_j = g^{x_j}$, $R_j = g^{r_j}$, $T_j = g^{t_j}$, then broadcast message $M_j^1 = (TID_{RSU_j}, E_{SK_{RSU_j}}(X_j, R_j, T_j), \sigma_{SK_{RSU_j}}(\cdot))$.

Step 2: $RSU_i$ receives all the message $\{M_j^1\}_{j \in \{1,\cdots,n\}, j \neq i}$, decodes it and gets $\{X_j, R_j, T_j\}$, then proceeds as following types:

1. Compute $Y_i^L = X_{i-1}^{x_i}, Y_i^R = X_{i+1}^{x_i}$, $Y_i = Y_i^R / Y_i^L$.

2. Compute $v_i = H(Y_i^L \| Y_i^R \| X_1 \| X_2 \| \cdots \| X_n \| pid)$, here $pid = h(TID_{RSU_1} \| TID_{RSU_2} \| \cdots \| TID_{RSU_n})$, and using $SK_{RSU_i} = \xi_i$ compute $s_i = r_i - v_i \cdot \xi_i$.

3. Using $r_i$ compute $T_{i,j} = T_j^{r_i} (j = \{1, \cdots, i-1, i+1, \cdots n\})$.

4. Broadcast $M_i^2 = (TID_{RSU_i}, E_{SK_{RSU_i}}(Y_i, s_i, T_{i,1}, \cdots, T_{i,i-1}, T_{i,i+1}, \cdots, T_{i,n}), \sigma_{SK_{RSU_i}}(\cdot))$ Session key generation $RSU_j$ receives all message $\{M_i^2\}_{j \in (1,\cdots,n), j \neq i}$, proceeds as follows:

1. Verify $T_{i,j} = R_i^{t_j} (i = \{1, \cdots, j-1, j+1, \cdots n\})$ holds or not. If it holds, continues, else, stop.

2. Compute $\hat{Y}^R_{j+1} = Y_{j+1} \cdot Y^R_j, \hat{Y}^R_{j+2} = Y_{j+2} \cdot \hat{Y}^R_{j+1}, \cdots, \hat{Y}^R_{j+(n-1)} = Y_{j+(n-1)} \cdot \hat{Y}^R_{j+(n-2)}$ and confirm $Y^L_j = \hat{Y}^R_{j+(n-1)}$. If it holds to all $RSU_i$, continues, else, cease.

3. All $RSU_i$ compute $\hat{v}_i = H(\hat{Y}^R_{i-1} \| \hat{Y}^R_i \| X_1 \| \cdots \| X_n \| pid)$, verify. If it holds to all $RSU_i$, continues, else, cease.

4. Compute $sk = \hat{Y}^R_1 \cdot \hat{Y}^R_2 \cdots \hat{Y}^R_n = g^{x_1 x_2 + x_2 x_3 + \cdots + x_n x_1}$.

4.3 The authentication between RSU and vehicles

With the RSU as the center, which has a wide range of communication and strong computing power, establish the group according to the geographic area on the road. After the vehicle get into the range of RSU, vehicles need to verify identity and negotiating key to join the group. And then vehicles can communicate with nearby vehicles and RSU. In this paper, the authentication is going on between RSU and vehicles without TA participating in it. At the same time, the legal vehicle will be authenticated with few times by using group key transmission mechanism. The authentication between $RSU_i$ and vehicle $V_i$ is shown in Fig. 2

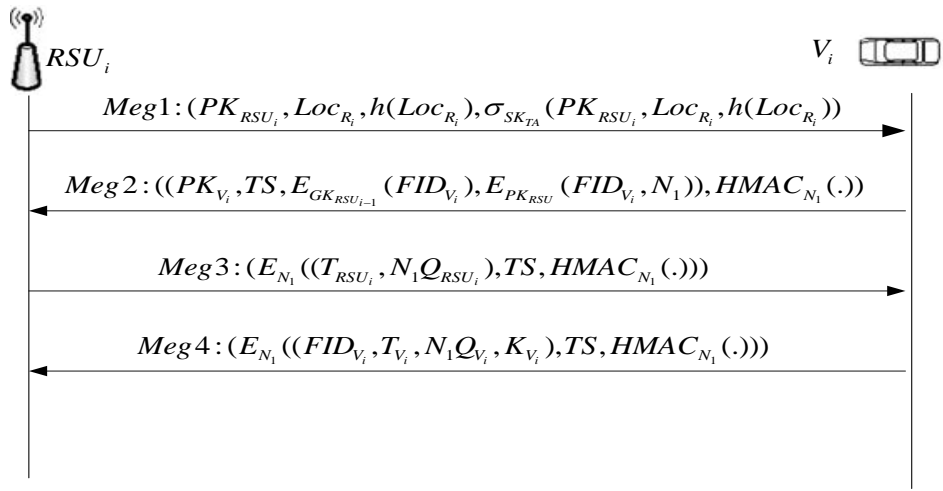

**Fig. 2.** The authentication between $RSU_i$ and $V_i$

Step 1: Each RSU store the signing message $\sigma_{SK_{TA}}(PK_{RSU_i}, Loc_{RSU_i}, h(Loc_{RSU_i}))$ which promulgate by TA. Here $h(Loc_{RSU_i})$ is hash operation on location information of $RSU_i$. $RSU_i$ Periodically broadcast message $Meg1:(PK_{RSU_i}, Loc_{RSU_i}, h(Loc_{RSU_i}), \sigma_{SK_{TA}}(PK_{RSU_i}, Loc_{RSU_i}, h(Loc_{RSU_i})))$.

Step 2: When it gets into the communication scope of $RSU_i$, vehicle $V_i$ received $Meg1$. $V_i$ gets $PK_{RSU_i}, Loc_{RSU_i}, h(Loc_{RSU_i})$, verifies $\sigma_{SK_{TA}}(PK_{RSU_i}, Loc_{RSU_i}, h(Loc_{RSU_i}))$. If it corrects, compute $h'(Loc_{R_i})$ by using $Loc_{RSU_i}$. If formula (1) holds, $V_i$ completes the authentication of $RSU_i$, else, discards the message.

$$h(Loc_{R_i}) \stackrel{?}{=} h'(Loc_{R_i}) \quad (1)$$

Step 3: If vehicle $V_i$ completed the step 2, then it chooses a random number $N_1$ and sends message $Meg2:((PK_{V_i}, TS, E_{GK_{RSU_{i-1}}}(FID_{V_i}), E_{PK_{RSU}}(FID_{V_i}, N_1)), HMAC_{N_1}(\cdot))$ to $RSU_i$.

Step 4: $RSU_i$ receives the message $Meg2$ verify time stamp $TS$ and calculates $\Delta t = CT - TS$, here $CT$ is current time. If $\Delta t$ meets the network delay, finishes the $TS$ verification, else discards the message. By using private key to decode $E_{PK_{RSU}}(FID_{V_i}, N_1)$, $RSU_i$ obtains $N_1, FID_{V_i}$. Then calculates $HMAC_{N_1}'(\cdot)$ and compares with $HMAC_{N_1}(\cdot)$, if them unequal, discards the message, else fetch the stored group key $GK$ which be transfer from the nearby RSU. Then, gets $FID'_{V_i}$ from $E_{GK_{RSU_{i\pm1}}}(FID_i)$ to compare $FID'_{V_i}$ and $FID_{V_i}$. If they equal, it explains that $V_i$ has been authenticated by the nearby RSU. So, the authentication is completed, executed the negotiation of the group key. Else $RSU_i$ chooses a random number $\alpha \in Z_q^*$ and computes $T_{RSU_i} = \alpha P$, then sends $Meg3:(E_{N_1}((T_{RSU_i}, N_1 Q_{RSU_i}), HMAC_{N_1}(\cdot)))$ to $V_i$.

Step 5: $V_i$ received $Meg3$, gets $T_{RSU_i}$ and verify $HMAC_{N_1}(\cdot)$. Then $V_i$ chooses a random number $\beta \in Z_q^*$, computes $T_{V_i} = \beta P$, $K_{V_i} = e(\beta N_1 Q_{RSU_i}, PK_{TA})e(N_1 s_{V_i}, T_{RSU_i})$, and sends $Meg4$: $(E_{N_1}((FID_{V_i}, T_{V_i}, N_1 Q_{V_i}, K_{V_i}), HMAC_{N_1}(\cdot)))$ to $RSU_i$.

Step 6: $RSU_i$ receives the $Meg4$, obtains $FID_{V_i}, T_{V_i}, N_1 Q_{V_i}, K_{V_i}$ from it, verify $HMAC_{N_1}(\cdot)$ and calculates $K_{RSU_i} = e(\alpha N_1 Q_{V_i}, PK_{TA})e(N_1 s_{RSU_i}, T_{V_i})$. If formula (2) holds, $RSU_i$ finishes the authentication to $V_i$, else, discards the message.

$$e(\beta N_1 Q_{RSU_i}, PK_{TA})e(N_1 s_{V_i}, T_{RSU_i}) \overset{?}{=} e(\alpha N_1 Q_{V_i}, PK_{TA})e(N_1 s_{RSU_i}, T_{V_i}) \qquad (2)$$

4.4 The negotiation and update of the group key

1) The negotiation of the group key

After the vehicle has been completed the authentication, it requires to communicate with RSU and other vehicles. Therefore, the vehicle need to join in the group centered on RSU and negotiate the group key ($GK$). The negotiation between $RSU_i$ and $V_i$ is shown in Fig. 3

Step 1: Vehicle $V_i$ chooses a random number $\lambda_i \in Z_q^*$, computes $g^{\lambda_i}$ and send $Pag1$: $(E_{N_1}((FID_i, g^{\lambda_i}), HMAC_{N_1}(\cdot)))$ to $RSU_i$.

Step 2: $RSU_i$ receives the message $Pag1$, get $FID_i, g^{\lambda_i}$ from it. Then chooses a random number $\gamma \in Z_q^*$, computes $g^{\lambda_i \gamma}, \prod_{i=1}^{n} g^{\lambda_i \gamma}$, new group key $GK = g^\gamma * \prod_{i=1}^{n} g^{\lambda_i \gamma}$, and then sends $Pag2$: $(E_{N_1}((g^{\lambda_i \gamma}, \prod_{i=1}^{n} g^{\lambda_i \gamma}), HMAC_{N_1}(\cdot)))$ to $V_i$, broadcast $Pag3: (E_{GK'}((GK) \| MAC_{GK'}(\cdot)))$ to the primary group members. Here $GK'$ is the primary group key.

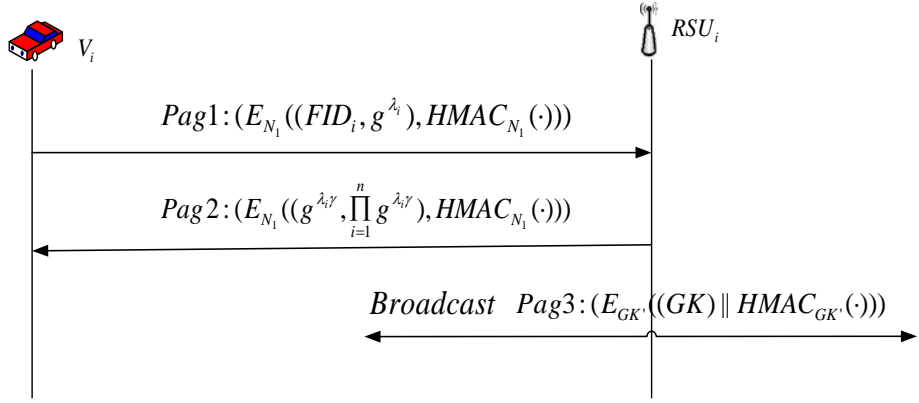

Fig. 3 The negotiation of the group key between $RSU_i$ and $V_i$

Step 3: $V_i$ receives $Pag2$, decodes it to get $g^{\lambda_i\gamma}, \prod_{i=1}^{n} g^{\lambda_i\gamma}$, computes $g^\gamma = (g^{\lambda_i\gamma})^{-\lambda_i}$ and $GK = g^\gamma * \prod_{i=1}^{n} g^{\lambda_i\gamma}$. Here $V_i$ has joined in the group which formed by $RSU_i$.

The group key transmission scheme: $RSU_i$ sends $E_{sk}(GK)$ to the nearby RSU through the wire communication when $GK$ has updated.

2) The update of the group key

In order to guarantee the communication of primary group members do not be affect and the leaved vehicle cannot communicate with the group members. The group key require to be update when vehicles join or leave the group.

Step 1: When vehicle $V_j$ has leaved the scope of $RSU_i$, $RSU_i$ chooses a random number $\gamma \in Z_q^*$, computes $g^{\lambda_i\gamma}$ of every group members except $V_j$ and their sum $\prod_{1}^{j-1}(\prod_{i+1}^{n} g^{\lambda_i\gamma})$. Broadcast

$Bm1: (E_{GK'}((g^{\lambda_1\gamma}, FID_{V_1}) \cdots (g^{\lambda_{j-1}\gamma}, FID_{V_{j-1}}), (g^{\lambda_{j+1}\gamma}, FID_{V_{j+1}}) \cdots (g^{\lambda_n\gamma}, FID_{V_n}), \prod_{1}^{j-1}(\prod_{i+1}^{n} g^{\lambda_i\gamma}), HMAC_{GK'}(\cdot)))$

Step 2: The group member $V_i$ receives $Bm1$. According gets $g^{\lambda_i\gamma}$ and $\prod_{1}^{j-1}(\prod_{i+1}^{n} g^{\lambda_i\gamma})$ by using the primary group key $GK'$ to decode it and according to $FID_{V_i}$. Then computes $g^\gamma = (g^{\lambda_i\gamma})^{-\lambda_i}$ and the new group key $GK = g^\gamma * \prod_{1}^{j-1}(\prod_{j+1}^{n} g^{\lambda_i\gamma})$

## 4.5 The communication among the group

1) The broadcast communication

Vehicle $V_i$ broadcast $E_{GK}(m, FID_{V_i}, HMAC_{GK}(\cdot))$ if it wants to broadcast message $m$ to other vehicles.

2) Communication between vehicles and RSU

When $V_i$ wants to send a message $m$ to $RSU_i$, sends $((FID_{V_i}, E_{N_1}(m)), HMAC_{N_1}(\cdot))$ to $RSU_i$.

3) Communication between vehicles

If $V_i$ needs one to one communication with $V_j$, they require to execute this communication process as shown in Fig. 4

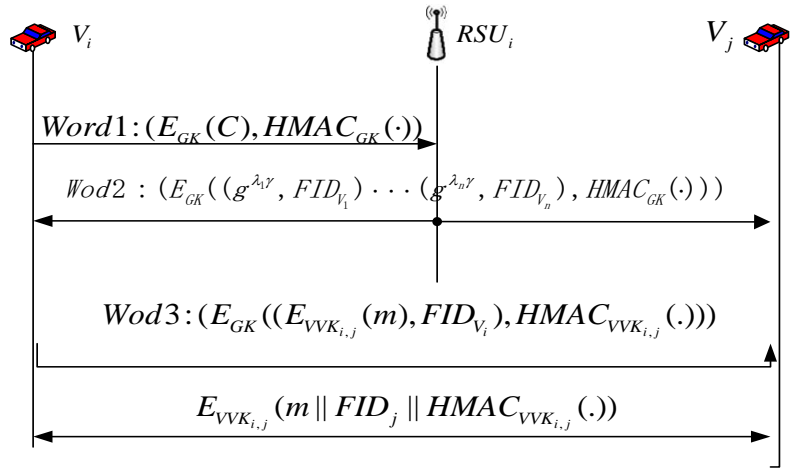

Fig. 4 The communication between $V_i$ and $V_j$

Step 1: $V_i$ sends $Wod1:(E_{GK}((C), HMAC_{GK}(\cdot)))$ to $RSU_i$, here $C$ is a fixed value, it explains the request of one to one communication.

Step 2: $RSU_i$ receives $Wod1$, obtains $C$ by using $GK$ decode it and according to $C$ broadcast the message $Wod2:(E_{GK}((g^{\lambda_1\gamma}, FID_{V_1})\cdots(g^{\lambda_n\gamma}, FID_{V_n}), HMAC_{GK}(\cdot)))$.

Step 3: $V_i$ receives $Word2$ verifies $HMAC_{GK}(\cdot)$. Then obtains $g^{\lambda_i\gamma}$ according to $FID_{V_j}$, computes $VVK_{i,j} = g^{\lambda_i\lambda_j\gamma}$, and then sends $Wod3:(E_{GK}((_{VVK_{i,j}}(m), FID_i, HMAC_{VVK_{i,j}}(\cdot)))$ to $V_j$.

Step 4: $V_j$ receives $W_{od2}$ and $W_{od3}$, obtains $g^{\lambda_i \gamma}$, computes $VVK_{i,j} = g^{\lambda_i \lambda_j \gamma}$ and verifies $HMAC_{VVK_{i,j}}(\cdot)$, if it is true, receives the message $m$.

**5. Analysis Of The Proposed Scheme**

In this section, we will analysis the scheme from security requirements which have been referred to in Introduction.

1) Message confidentiality

For the authentication and negotiation in section 4.4, message is protected by the shared secret key $N_1$, the cipher text under symmetric encryption cannot leak any information about message. The broadcast in section 4.5, confidentiality is protected by the shared secret group key and the confidentiality of one to one communication is protect by the shared secret key $VVK_{i,j}$ between the vehicles.

2) Message integrity and authentication

This scheme use $E_k(m)$ and $HMAC_k(m)$ to realize the message integrity and confidentiality of message $m$. Now show that the attacker cannot generate a valid signature.

Have an assumption that the shared key $k$ is kept secret, then a node message cannot be forged by the attacker, and the scheme is secure against existential forgery, adaptive chosen message attack under random oracle model. First, consider the Game between the challenger and the attacker.

Setup: The challenger starts by giving the attacker a set system parameters.

Challenge: The challenger asks the attacker to pick a random message $m$ and sign it to generate $E_k(m)$ and $HMAC_k(m)$.

Guess: Finally, the attacker sends $E_k(m)$ and $HMAC_k(m)$ to the challenger.

The attacker's advantage is defined to be $E_k(m)$ and $HMAC_k(m)$ are valid signatures. The scheme is secure against existential forgery, adaptive chosen message attack if the attacker's advantage is negligible.

In section 4.3, $k = N_1$ is the shared key between $V_i$ and $RSU_i$. $V_i$ encodes $N_1$ by $PK_{RSU_i}$ and transfers it to $RSU_i$. However, only $RSU_i$ has $SK_{RSU_i}$ to decode it and get $N_1$. So, only $V_i$ and $RSU_i$ know the key $N_1$. In section 4.5 the shared key between $V_i$ and $V_j$ is $VVK_{i,j} = g^{\lambda_i \lambda_j \gamma}$. However, $\lambda_i, \lambda_j$ is respectively held by $V_i$ and $V_j$, it do not be transferred. According to the difficulty problem, it is known that only $V_i$ and $V_j$ can compute $VVK_{i,j}$

The above analysis shows that the attacker cannot obtain the shared secret key, and cannot forge any valid message. So, the attacker's advantage is negligible and our scheme is secure.

3) Identity privacy

The message which sent by $V_i$ only contained pseudonym $FID_{V_i}$, the public key $PK_{V_i}$ and $N_1 Q_{V_i}$. Vehicles will generate a new random $\alpha_i$ when it gets into the scope of $RSU$. So pseudonym $FID_{V_i} = TID_{V_i} \oplus (\alpha_i * PK_{TA})$ and the public key $PK_{V_i} = g^{\alpha_i}$ of one vehicle is diffident within the scope of different RSU. Therefore, the attacker cannot attack the vehicle by pseudonym and the public key. According to CDH problem which has been mention in section 3.2(2), the attacker cannot compute $Q_{V_i}$ from $N_1 Q_{V_i}$. However, $N_1 Q_{V_i}$ is also different in the scope of different RSU. Therefore, the attacker also cannot attack the vehicle by $N_1 Q_{V_i}$.

4) Non-repudiation

To avoid the vehicle deny ever sent the message which led to the accident, TA should be able to reveal the true identity of the sender from the message. In this scheme, TA can get the pseudonym from the message, and get the public key of the sender with RSU assist. Then, TA uses its private key, the public key $PK_{V_i}$ of the sender and the pseudonym to have a computation as following:

$$TID_{V_i} = FID_{V_i} \oplus H_1(s * PK_{V_i}) = TID_{V_i} \oplus H_1(\lambda_i * PK_{V_i}) \oplus H_1(s * PK_{V_i}) = TID_{V_i}$$

Finally, TA obtain the true identity $TID_{V_i}$ of sender, and in addition to TA and the vehicle itself, the other participants cannot know $TID_{V_i}$. Therefore, our scheme is non-repudiation.

5) Forward-security and backward-security

When the vehicle $V_j$ has left the scope of $RSU_i$, $RSU_i$ random selects $l \in Z_q^*$ and computes the new group key is $GK = g^l * \prod_{1}^{j-1}(\prod_{j+1}^{n} g^{r_l})$. The vehicle $V_j$ only know $(g^{\lambda_1\gamma}, g^{\lambda_2\gamma} \cdots g^{\lambda_{j-1}\gamma}, g^{\lambda_{j+1}\gamma}, \cdots, g^{\lambda_n\gamma}, \prod_{1}^{j-1}(\prod_{j+1}^{n} g^{\lambda_i\gamma}))$, but could not know the new $g^{\lambda_j\gamma}$. According to the DDH problem, $V_j$ cannot compute $g^\gamma$ and the new group key. Therefore, our scheme is forward-secure. When the vehicle $V_i$ joined in the group, it received the message by RSU to update the group key. However the message is encode by the previous group key, vehicle cannot decode it and get the previous group key. So, our scheme is backward-secure.

## 6. Performance Evaluation

In this section, we compare with some related work from verification delay and transmission overheard, then have a simulation on message delay by NS2.34.

6.1 Verification delay

The experiment is running on an Intel Pentium IV 3.0 GHZ machine proposed in reference[14]. According to [6], the following results are obtained: the time of a pairing operation $T_{par}$ is 4.5ms, the time of performing one point multiplication over an elliptic curve $T_{mul}$ is 0.6ms and the time of a MapToPoint hash operation $T_{mtp}$ is 0.6ms. The computation cost of the message certification mainly focus on the above three parameters, any other operations are not considered, such as each HMAC operation is assumed to take 0.006ms. TABLE 2 shows the comparison of verification delay of other schemes.

TABLE 2 COMPARISION OF VERIFICATION DELAY

| Scheme | Complete n verification | |
|---|---|---|
| | OBU | RSU |
| IBV[6] | $5nT_{mul} + 2nT_{mtp}$ | $(n+1)T_{mul} + 3T_{par} + nT_{mtp}$ |
| ECPP[16] | $4nT_{mul} + nT_{par}$ | $2nT_{mul} + 3nT_{par}$ |
| RMAKA[17] | $4nT_{mul}$ | $4nT_{mul}$ |
| ACP[15] | $nT_{mul}$ | $3T_{par} + (2n+1)T_{mul}$ |
| Our scheme | $nT_{mtp} + 5nT_{mul} + 2nT_{par}$ | $4nT_{mul} + 2nT_{par}$ |

Fig. 5 illustrates that the verification delay radio compared with others scheme for RSU verify the vehicles when the illegal vehicles is 5%. From the Fig. 5, with the increase of certified vehicles, the ratio of ACP, IBV are on the rise, but they are less than 1. Therefore, the verification delay of our scheme is less than them. And compared with the ECPP, RMAKA, the verification speed of our scheme is about 94% faster than that of ECPP, and is about 75% faster than that of RMAKA.

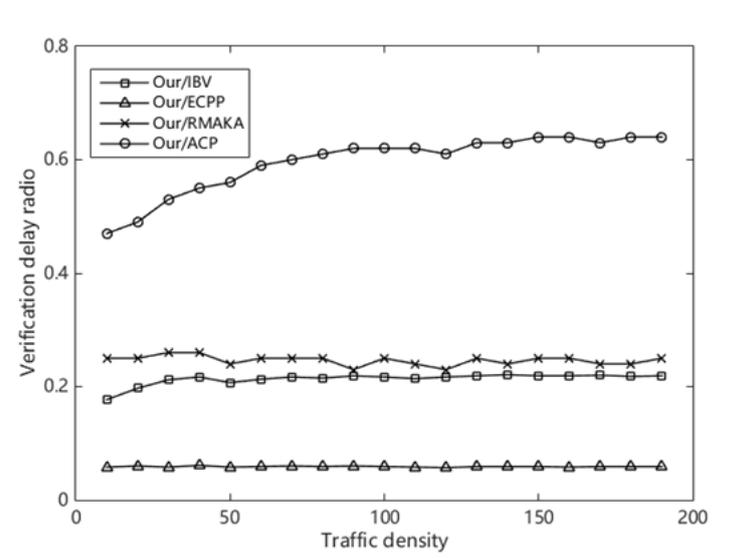

Fig 5 Traffic density and Verification delay radio

6.2 Transmission overhead

**TABLE 3.** COMPARISONS OF TRANSMISSION OVERHEAD

| Scheme | Send $n$ messages (bytes) |
|---|---|
| | $OBU \rightarrow RSU$ |
| RMAKA[17] | 167n |
| ABAKA[19] | 84n |
| ARGBV[20] | 63n |
| Our scheme | 58n |

According to reference[18], each vehicle send message every 300ms. In this paper, the length of pseudonym is 42bytes and HMAC is 16bytes. Fig. 6 and TABLE 3 illustrates that the transmission overhead of RMAKA, ABAKA, ARGBV and our scheme. From the Fig. 6, we can see the transmission overhead of us is least.

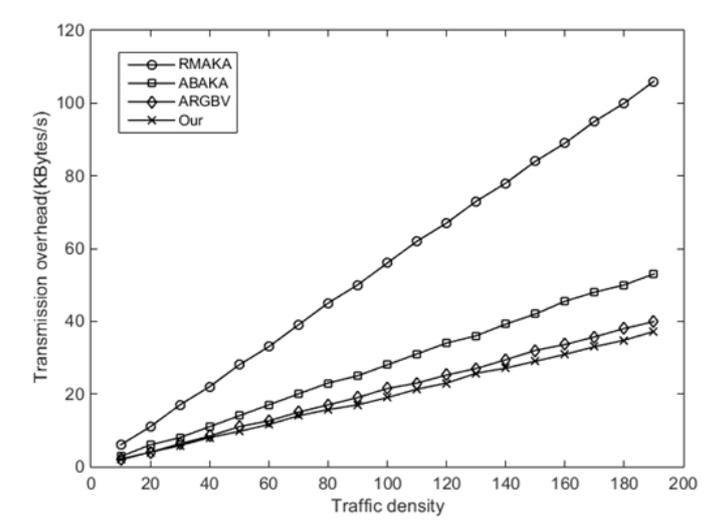

**Fig. 6.** Traffic density and Transmission overhead

6.3 Simulation

1) The simulation parameter

The main simulation parameters are list in TABLE 4

**TABLE 4.** SIMULIATION PARAMETERS

| Simulation parameters | Value |
| --- | --- |
| Road length | 1000 m |
| Simulation time | 20 s |
| Message size | 200 bytes |
| Broadcast interval | 300 ms |
| Interval variance | 0.05 s |
| Communication Range of RSU and Vehicles | 600 m |
| Communication Range of Vehicles | 300 m |
| Bandwidth | 6 Mbps |

2) Message delay

We define the average delay of a message as [15]:

$$Delay = \frac{1}{N}\sum_{n=1}^{N}\frac{1}{M_i}\sum_{m=1}^{M}(T_{Creat}^{n-m} + T_{Transmission}^{n-m-k} + T_{Verify}^{n-m-k})$$

Here $N$ is the number of the vehicles, $M_i$ is the number of messages sent by vehicle $V_i$, $T_{Creat}^{n-m}$ is the time that the Vehicle or RSU create the message, $T_{Transmission}^{n-m-k}$ is the transmission time that entity n sent message to entity $k$, $T_{Verify}^{n-m-k}$ is the time that entity $k$ verify the message from entity n.

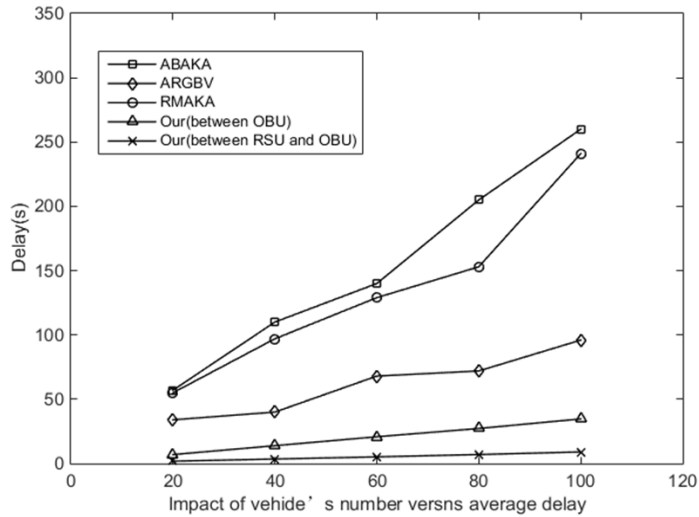

Fig. 7 The message delay and traffic density

Fig. 7 illustrates the impact of vehicle's number versus average delay. In this paper, the communication of us divided on two aspects. One is between OBU, and another is between RSU and OBU. Fig. 7 shows that the message delay also increase with the increase of vehicle number. The message delay of ABAKA, ARGBV and RMAKA are bigger than us. So, with the increase of vehicle quantity, the message delay of our scheme is least.

## 7. Conclusions

In this paper, we proposes a self-authenticated deniable efficient group key agreement scheme in VANET. The features of the scheme as following: (1) Employ the no certificate public key system, the authentication process between vehicles and RSU has no authentication center participated in, avoid the time delay problem of TA certificate to speed up the certification. (2) Reduce the frequency of legal vehicles' certification and avoid tracking legal vehicles through RSU by using the deniable group key transmission scheme. (3) In order to alleviate the workload of the group leader and eliminate the possible single point of failure problem by using key negotiation instead of the group leader distribute group key. In the future, we will have a further research on how the vehicles build the group for secure communication by themselves on the way of no public infrastructure.

**Acknowledgement**

This research is supported by the following funds: Natural Science Foundation of China (61300229), Six Talent Peaks Project of Jiangsu Province (DZXX-012), Natural science fund for colleges and universities in Jiangsu Province (Grant No.12KJD580002) and Traffic information fund of ministry of communications of China (Grant No. 2013-364-863-900)